\documentstyle[12pt,aasms4]{article}
\raggedright

\tighten

\begin{document}
\title{Detection and Mapping of Decoupled Stellar and Ionized Gas Structures
in the
Ultraluminous Infrared Galaxy IRAS 12112+0305$^1$}

\vspace{0.5in}

\author{Luis Colina\altaffilmark{2},
Santiago Arribas\altaffilmark{3,4},
Kirk D. Borne\altaffilmark{5}, \&
Ana Monreal\altaffilmark{3}}

\vspace{0.25in}

\scriptsize
\affil{(2) \em Instituto de F\'{\i}sica de Cantabria (CSIC-UC), Facultad
 de Ciencias, Avda. de Los Castros S/N, 39005 Santander, Spain
(colina@ifca.unican.es)}
\affil{(3) \em Instituto de Astrof\'{\i}sica de Canarias, V\'{\i}a L\'actea
S/N,
38200 La Laguna, Tenerife, Spain (sam@ll.iac.es)}
\affil{(4) \em Consejo Superior de Investigaciones Cient\'{\i}ficas (CSIC)}
\affil{(5) \em Raytheon Information Technology and Sciences Services, NASA
Goddard
Space Flight Center, Greenbelt, MD20771, USA (borne@rings.gsfc.nasa.gov)}

\vspace{1in}
\footnote{Based on observations with WHT operated on the\
 island of La Palma by the ING in the Spanish Observatorio del Roque de los
 Muchachos of the Instituto de Astrof\'{\i}sica de Canarias.
 Based also on observations with the NASA/ESA Hubble Space Telescope,
 obtained at the Space Telescope Science Institute, which is operated by
 the Association of Universities for Research in Astronomy, Inc. under
 NASA contract No.  NAS5-26555.}

\normalsize

\begin{abstract}

Integral field optical spectroscopy with the INTEGRAL fiber-fed
system and {\it{HST}} optical imaging are
used to map the complex stellar and warm ionized gas structure in
the ultraluminous infrared galaxy
IRAS 12112+0305.
Images reconstructed from wavelength-delimited extractions of the
integral field spectra
reveal that the observed ionized gas distribution
is decoupled from the stellar main body of the galaxy, with the
dominant continuum
and emission-line regions separated by 
projected distances of up to 7.5 kpc.
The two optical nuclei are detected as apparently faint 
emission-line regions, and their optical properties are consistent with
being dust-enshrouded
weak-[O\thinspace I] LINERs.
The brightest emission-line region
is associated with a faint ($m_{\rm{I}}$= 20.4), giant
H\thinspace II region of 600 pc diameter, where a young ($\sim$ 5 Myr) massive
cluster of about 2 $\times$ 10$^7$ $M_{\odot}$ dominates the ionization.
Internal reddening towards the line-emitting regions and
the optical nuclei ranges from 1 to 8 magnitudes, in the visual. 
Taken into account the reddening, the overall star formation
in IRAS 12112+0305 is dominated by starbursts associated with the two
nuclei and corresponding to
a star formation rate of 80 $M_{\odot}$ yr$^{-1}$.

\end{abstract}


\section {INTRODUCTION}

Ultraluminous infrared galaxies (ULIRGs), with bolometric luminosities
$L_{\rm{bol}} \approx$
$L_{\rm{IR}} \geq 10^{12}$ $L_{\odot}$,
are the most luminous galaxies in the local Universe. ULIRGs show signs of
strong interactions and
mergers (Melnick \& Mirabel 1990, Clements et al. 1996, Borne et al. 2000a),
 or even multiple collisions (Borne et al. 2000b), and 
they have large
amounts
of gas and dust that significantly obscure the nuclear ionizing sources
(Sanders \&
Mirabel 1996). ULIRGs appear to be the
low-redshift analogs of the high-redshift submillimeter galaxies (SCUBA
sources),
which are responsible for the bulk of the far-infrared background radiation
(see Sanders 1999 for a review).

Mid-infrared spectroscopy has shown that ULIRGs with
optical H\thinspace II-
and LINER-like spectra are starbursts dominated (Lutz et al. 1998; Genzel et al. 1998;
Lutz,
Veilleux, \& Genzel 1999). However, the increased
fraction of optically classified Seyfert galaxies among the
more luminous
ULIRGs (log$L_{\rm{IR}}
\geq 12.3$ $L_{\odot}$) is taken as evidence for the presence of 
dust-enshrouded quasars powering these galaxies (Veilleux,
Kim, \& Sanders 1999).

The ionization structure of ULIRGs is extended and rather complex, and
therefore
narrow-slit ($\leq$ 2$^{''}$) optical
spectroscopy along a given position angle cannot adequately indicate the
location,
nature, and interplay of the different ionization
mechanisms present in these galaxies. The spectral
classification of some ULIRGs are already known
to depend on the size of the aperture (Kim, Veilleux, \& Sanders 1998).
Two-dimensional optical spectroscopy using fiber-fed spectrographs
is an ideal technique to study the complex stellar and ionized gas structure
of the
nuclear regions of galaxies
in general (Arribas \& Mediavilla 2000), and of ULIRGs in particular
since it allows a simultaneous and complete mapping of the stellar
populations and the
ionized gas (e.g., see Colina, Arribas, \& Borne 1999).

IRAS 12112+0305 is a low
redshift
(z= 0.0723) ULIRG (log$L_{\rm{IR}}$= 12.34 $L_{\odot}$), classified in the
optical as
a LINER (Kim et al. 1998), and with no evidence for a hidden
broad-line
region (Veilleux, Sanders, \& Kim 1999). Mid-infrared diagnostics
classify this galaxy as starburst-dominated (Genzel et al. 1998).
Near-infrared ground-based images
revealed the presence of two nuclei (Carico et al. 1990), while {\it{HST}} 
optical
and near-infrared images show a well defined compact point-like nucleus,
as well as several much fainter condensations distributed in an arc-like
structure (Borne et al. 2000a; Scoville et al. 2000).

In this letter we highlight new results for IRAS 12112+0305 obtained
using
integral field optical spectroscopy and {\it{HST}} imaging. Throughout the paper
a Hubble constant of
70 km s$^{-1}$ Mpc$^{-1}$ is assumed.

\section{OBSERVATIONS AND DATA REDUCTION}

Integral field spectroscopy of IRAS 12112+0305 was obtained with the
INTEGRAL system
(Arribas et al. 1998), a fiber-fed
spectrograph mounted at the Naysmith No. 1 platform of the 4.2m William
Herschel Telescope. The bundle of fibers consisted
of 219 fibers,
each 0.9$^{''}$ in diameter and
covering a 16.5$^{''}$ $\times$ 12.3$^{''}$ field-of-view. The spectra
were taken using
a 600 line/mm grating, covering the
5000$-$7900\AA~ range, with an effective resolution of 4.8\AA. The total
integration time
was 7500 sec, split into five
separate integrations of  1500 sec each, with seeing 
$\approx 1.0^{''}$.

The reduction has been done following the standard procedures applied to 
spectra obtained with two-dimensional fiber spectrographs (Arribas et al. 1997
and references therein).
The results are presented in Figure~1, together with the {\it{HST}} 
$I$-band
image
(Borne et al. 2000a) for comparison.

The derived astrophysical properties for the main continuum and
line-emitting 
regions are presented in Table~1. Columns 2 and 3 give the relative
positions. Column
4 presents the internal reddening values derived
from the H$\alpha$/H$\beta$ ratio, assuming case B recombination. Column 5
gives
the apparent magnitude within an aperture of 0.5$^{''}$ radius using the
{\it{HST}} $I$-band image, and column 6 gives the corresponding absolute 
magnitude
after
internal reddening correction assuming $A_{\rm{I}}$= 1.494 $\times$ E(B$-$V).
Column 7
shows the observed H$\alpha$ flux obtained using an aperture
of 2.1$^{''}$ diameter, while column 8 gives the reddening-corrected
H$\alpha$
luminosity. The last four columns present the logarithm of the
reddening-corrected
emission line ratios and the corresponding activity classification.

\section{RESULTS AND DISCUSSION}

\subsection{Evidence for Decoupled Stellar and Ionized Gas Components}

The stellar main body of IRAS 12112+0305 is concentrated in three dominant
regions, each
separated from each other by about 2$^{''}-3^{''}$ (see the H$\alpha$ and
H$\beta$ continuum
images in Figure~1). Two of these regions (called N$_s$ and N$_n$
hereinafter; see Table~1 for
relative positions) located along position angle PA45, were already detected
in the near-infrared,
and associated with the nuclei of two galaxies involved in the final stages
of a merger (Carico et al. 1990). The
INTEGRAL-generated
continuum images extend the wavelength coverage towards the blue, clearly
showing the presence of a large differential extinction towards the southern
nucleus (N$_s$).
The third region (called R2 hereinafter) is located 3$^{''}$ north of
N$_s$ at position angle PA20.

The {\it{HST}} $I$-band image (Figure~1) shows N$_s$ as a high-surface brightness compact 
region
which coincides with the K-band point-like nucleus
(Scoville et al. 2000). Regions N$_n$ and R2 are made of
several fainter condensations distributed along an
arc-like structure of about 5.4$^{''}$ (i.e.,
8 kpc) extent and located 3$^{''}$ to 5$^{''}$ (i.e., 4.5 to 7.5 kpc)
north$-$northeast
of N$_s$.
The brightest of these condensations, located 
3$^{''}$ ($\approx$ 4.5 kpc) northeast
from N$_s$, is most likely associated with the northern nucleus detected in
the INTEGRAL-generated
continuum images (N$_n$) and in the near-infrared (Carico et al.
1990; Scoville et al.
2000). 

Although the overall
structure of the ionized gas resembles that of the stellar light
distribution,
the dominant line-emitting regions do not coincide with the nuclei
identified above
but, on the contrary, are decoupled from them (see [O\thinspace III] and 
H$\alpha$ maps
in Figure~1).
The brightest line-emitting region (called R1 hereinafter) is located 
5$^{''}$ ($\approx$ 7.5 kpc) 
east of N$_s$ along position angle PA80. This region is
associated with
a faint $I$-band continuum source ($m_I$= 20.4) outside the main body
of the galaxy (see {\it{HST}} $I$-band image in Figure~1). The second-brightest 
H$\alpha$ line-emitting region
is associated with region R2, and it is therefore 
composed of a filament
of faint continuum condensations that are also detected in the H$\beta$
and [O\thinspace III] maps
(Figure~1).
The optically dominant nucleus of the galaxy (N$_s$) is marginally detected 
in [O\thinspace III]
and appears as the faintest H$\alpha$ emission source. The arc-like
structure seen in the {\it{HST}} $I$-band image corresponds to
three well resolved H$\alpha$
line-emitting peaks, the faintest of which is associated with the northern
nucleus (N$_n$) detected in the optical and near-infrared (see discussion
above).

\subsection{Massive Dust-Enshrouded Starbursts as Nuclear Ionizing Sources}

The positions of the two apparently faint ionizing nuclei N$_n$ and N$_s$
coincide (within the {\it{HST}} absolute astrometry uncertainties 
of $\pm$0.7$^{''}-$1$^{''}$)
with the two bright compact starburst nuclei detected at radio frequencies
(Condon et al. 1991), thus favoring dust-enshrouded nuclear starbursts as the
energy
sources in these nuclei.

The ionization sources in N$_s$ and N$_n$ are highly obscured, with visual
extinctions
($A_{\rm{V}}$) of 8 and 3.5 magnitudes, respectively (see spectra in Figure 2). 
The two nuclei
appear almost equally bright with $I$-band absolute magnitudes of $-$18.9
(N$_s$) and
$-$18.5 (N$_n$). However, when reddening is taken into
account, N$_s$ is found to be ten times more luminous than N$_n$ (i.e. 
$M_I = -$22.8).
These magnitudes are within the range of those measured for the
nuclei of other luminous and ultraluminous infrared galaxies (Surace et al. 1998).

The line ratios of the ionization sources associated
with the nuclei N$_s$ and N$_n$ correspond to a mixture of
weak-[O\thinspace I] LINER and H\thinspace II region spectra (see Table~1).
The presence of weak-[O\thinspace I] LINERs in the nucleus of galaxies has been 
taken
as evidence for
ionization by hot stars in a high-metallicity environment (Filippenko \&
Terlevich 1992;
Shields 1992) or for ionization 
by a mixture of a low-luminosity AGN and hot stars (Ho et
al. 1993).
The integrated mid-infrared spectrum of IRAS 12112+0305 does not show any
evidence
for an AGN (Genzel et al 1998), further supporting the idea that
the ionizing
sources in the nuclei N$_s$ and N$_n$ are associated with dust-enshrouded
starbursts.
Although N$_s$  and N$_n$ are minor contributors to the observed H$\alpha$
emission,
the reddening-corrected flux emanating from these nuclei dominates
the overall H$\alpha$ luminosity with a value of 9.1 $\times$ 10$^{42}$ erg
s$^{-1}$
(Table~1). If the H$\alpha$ flux emitted by the two nuclei
were entirely due to stars, then the corresponding star formation rate would
amount to about
80 $M_{\odot}$ yr$^{-1}$ for a Salpeter initial mass function (IMF) with
mass limits of 0.1 and 100 $M_{\odot}$ (Leitherer et al. 1999).

\subsection{Brightest Line-Emitting Region: A Tidally-Induced Giant H\thinspace 
II
Region?}

The {\it{HST}} image (Figure~1) shows that the apparently more luminous 
line-emitting peak (R1) is
associated with a faint ($m_I$= 20.35) 
region characterized by a physical size of about 600 parsecs ($\approx$
0.4$^{''}$), an
$I$-band absolute magnitude of $-$17.7, an internal optical extinction of 
about one magnitude, and an H$\alpha$
luminosity of 8.7 $\times$ 10$^{40}$ erg s$^{-1}$ (see Table~1).  

The size, H$\alpha$ luminosity and 
emission line ratios are typical of
circumnuclear star-forming regions in nearby spirals  
(Gonzalez-Delgado \& P\'erez 1997; Planesas, Colina, \& P\'erez-Olea 1997),
and
of giant extragalactic H\thinspace II regions like NGC 5471 in M101 (Shields 
1990).
 The H$\alpha$ luminosity and the equivalent
widths of the H$\beta$ (74 $\pm$ 3\AA) and H$\alpha$ (450 $\pm$
10 \AA) emission lines correspond to that of a 5-Myr old ionizing cluster
of 2 $\times$ 10$^{7}$ $M_{\odot}$, assuming a Salpeter IMF with mass
limits
of 0.1 and 100 $M_{\odot}$ (Leitherer et al. 1999). The mass of the ionizing
cluster
represents only a small fraction ($\sim$ 3\%) of the dynamical mass of this
region (upper limit
of 7.5 $\times$ 10$^8$
$M_{\odot}$), calculated assuming virialization and an instrument-corrected
emission
line width equivalent to a velocity dispersion of 60 km s$^{-1}$ (derived
from the line-width measurements of the H$\beta$ and [O\thinspace III]5007\AA~ 
lines).

In summary, the derived properties of this region are characteristic of
young massive
H\thinspace II regions and could represent a case for a tidally-induced
giant extranuclear self-gravitating star-forming region, or even a
dwarf galaxy, decoupled from the much older stellar body of
the parent galaxies (Duc \& Mirabel 1994, 1998).

\section{SUMMARY}

Integral field optical spectroscopy and {\it{HST}} imaging 
have revealed the complex stellar and ionized gas structures in 
the ultraluminous infrared galaxy IRAS 12112+0305.
The two nuclei detected in the optical coincide with previously known radio
and near-infrared
nuclei, but appear as the apparently two faintest line-emitting regions and are
consistent with dust-enshrouded
(A$_V$= 3.5 and 8 mag) weak-[O\thinspace I] LINERs. However, their
reddening-corrected H$\alpha$ emission dominates the overall H$\alpha$
luminosity of the
system and, if associated with nuclear starbursts, would correspond to a
star formation of 80 $M_{\odot}$ yr$^{-1}$.

The observed structure of the ionized gas is decoupled from, and 
hence does not trace, the stellar light distribution. The brightest line-emitting
peak is 
associated with a faint ($m_I$= 20.4)
region located 7.5 kpc from the dominant optical nucleus (N$_s$), 
well outside
the main stellar body of the system. This region 
appears to be a recent (tidally induced?) star-forming region containing a 
young (5 Myr), massive (2
$\times$
10$^{7}$ $M_{\odot}$) cluster of stars.

The results presented here for IRAS 12112+0305 stress
the need for integral-field spectrographs in the study of the complex 
ionization and stellar light structure of ultraluminous infrared galaxies. 
Similar studies for high-redshift analogs
of ULIRGs (i.e., the SCUBA sources), and for other morphologically complex
high-redshift galaxies, will become feasible
when integral-field spectrographs on giant telescopes
become operational.

\acknowledgments
L. Colina thanks the Instituto de Astrof\'{\i}sica de Canarias for
its hospitality and financial support. K. Borne thanks Raytheon for
providing
financial support during his Sabbatical Leave.
Support for this work was provided by CICYT (Comisi\'on Interministerial de
Ciencia y Tecnolog\'{\i}a) through grants numbers PB98-0340-C02-01 and
PB98-0340-C02-02, and by
NASA through grant number GO-06346.01-95A from the Space
Telescope Science Institute, which is operated by
AURA, Inc., under NASA contract NAS5-26555.

\newpage

\scriptsize

\begin{deluxetable}{cccccccccccc}
\tablecaption{Properties of the Emission Line Regions}
\tablehead{
Region & $\bigtriangleup\alpha$\tablenotemark{a} &
$\bigtriangleup\delta$\tablenotemark{a}
& E(B$-$V) &
 m$_I$\tablenotemark{b} & M$_I$ 
& F$_{obs}$(H$\alpha$)\tablenotemark{c} 
& L(H$\alpha$)\tablenotemark{d}
& [O\thinspace III]/H$\beta$ 
& [O\thinspace I]/H$\alpha$\tablenotemark{e} 
& [N\thinspace II]/H$\alpha$\tablenotemark{e}
& [S\thinspace II]/H$\alpha$\tablenotemark{e} \\
(1) & (2) & (3) & (4) & (5) & (6) & (7) & (8) & (9) & (10) & (11) & (12)
\\ }
\startdata
 N$_s$ & $-$2.0 & $-$2.2  & 2.62 & 18.61   & $-$22.8 & 1.75 & 83.9 & 0.25 &
$-$0.88
       & $-$0.15 & $-$0.50 \nl
       &  &  &      &         &         &      &      &      & (w-L)   &
(L)    &  (H/L) \nl
 N$_n$ & 0.0 & 0.0  & 1.13 & 18.96   & $-$20.2 & 4.43 & 6.9  & 0.41 &
$-$0.85 & $-$0.31 & $-$0.31 \nl
       &  &  &      &         &         &      &      &      & (w-L)   &
(H/L)   &  (L)  \nl
 R1    & 2.7 & $-$1.5 & 0.31 & 20.35   & $-$17.7 & 3.67 & 0.9  & 0.25 &
$-$1.35
       & $-$0.66 & $-$0.63 \nl
       &  &  &      &         &         &      &      &      &  (H)    &
(H)    &  (H)  \nl
 R2    & $-$1.5 & 0.6 & 1.11 & 19.16   & $-$20.0 & 4.40 & 6.5  & $-$0.02 &
$-$1.05 & $-$0.41
       & $-$0.34 \nl
       &  &  &      &         &         &      &      &      & (w-L) &  (H)
&  (L)  \nl
\tablenotetext{a}{Relative positions in seconds of arc as measured from the WFPC2 I-band
  image (see Figure~1).}
\tablenotetext{b}{Uncertainties of $\pm$0.05 mags.}
\tablenotetext{c}{Flux in units of 10$^{-15}$ erg s$^{-1}$ cm$^{-2}$
A$^{-1}$. Uncertainties of
10\%$-$15\%.}
\tablenotetext{d}{Reddening corrected luminosity in units of 10$^{41}$ erg s$^{-1}$.}
\tablenotetext{e}{L: LINER; H: H\thinspace II region; w-L: 
weak-[O\thinspace I] LINER with
 I([O\thinspace I])/I(H$\alpha$) $<$ 1/6; H/L: boundary of 
H\thinspace II and LINER.}
\enddata
\end{deluxetable}

\normalsize

\newpage

\figcaption{INTEGRAL images of the ionized gas and stellar light
distribution in the
central regions of IRAS 12112+0305 as traced by different emission lines
(H$\beta$,
[O\thinspace III]5007\AA, and H$\alpha$), and by continuum windows close 
to H$\beta$
and H$\alpha$.
The high-resolution {\it{HST}} $I$-band image is also shown 
for
comparison.
The ionized gas morphology does not follow the stellar light distribution.
The southern
nucleus (N$_s$), associated with the point-like $I$-band region seen
in the {\it{HST}}
image, appears as a low-surface brightness emission peak
in H$\alpha$. The emission line gas is dominated by a region (R$_1$) located 
5$^{''}$ ($\approx$
7.5 kpc) east-northeast of N$_s$ (see text for details).}

\figcaption{Plots showing spectra of the main line-emitting regions (R1 \& R2),
 and of the northern and southern nuclei (N$_n$ and N$_s$, respectively). 
 Note that the horizontal axis does not represent a continuous range in wavelength, 
but the plots present two subsets of the full INTEGRAL spectrum
corresponding to the H$\beta$-[O\thinspace III] and 
H$\alpha$-[S\thinspace II] spectral regions,
respectively.}


\newpage

\begin{thebibliography}{}



\bibitem[Arribas {\it{et al.}}~1998]{ARR98} Arribas, S., et al. 1998 SPIE, 3355, 821.
\bibitem[Arribas \& Mediavilla 2000]{ARR00} Arribas, S., Mediavilla, E. 2000
    in  'Imaging the Universe in Three Dimensions:
    Astrophysics with Advanced Multi-wavelength Imagin Devices',  ASP, Conf.
    Ser. (in press).
\bibitem[Arribas {\it{et al.}}~1997]{ARR97} Arribas, S., Mediavilla, E., 
     Garc\'{\i}a-Lorenzo, B. \& del Burgo, C. 1997, ApJ 490, 227
\bibitem[Borne et al. 2000]{BBCL} Borne, K. D., Bushouse, H., Colina, L., \&
Lucas, R. A.
         2000a, in prep
\bibitem[Borne et al. 2000]{BBLC} Borne, K. D., Bushouse, H., Lucas, R. A., \&
Colina, L.
         2000b, ApJ 529, L77
\bibitem[Carico et al. 1990]{CAR90} Carico, D.P., Graham, J.R., Matthews,
K., Wilson,
         T.D., Soifer, B.T., Neugebauer, G., \& Sanders, D.B. 1990, ApJ 349,
L39
\bibitem [Clements et al. 1996][CL96] Clements, D.L., Sutherland, W.J., McMahon, R.G.
         \& Saunders, W. 1996, MNRAS 279, 477
\bibitem[Colina, Arribas \& Borne 1999]{CAB} Colina, L., Arribas, S., \&
Borne, K. D.
         1999, ApJ 527, L13
\bibitem[Condon et al. 1991]{CHYT} Condon, J.J., Huang, Z.P., Yin, Q.F., \&
Thuan, T.X.
         1991, ApJ 378, 65
\bibitem[Duc \& Mirabel 1994]{DM94} Duc, P.A., \& Mirabel, I.F. 1994,
         A\&A 289, 83
\bibitem[Duc \& Mirabel 1998]{DM98}Duc, P.A., \& Mirabel, I.F. 1998
         A\&A 333, 813
\bibitem[Filippenko \& Terlevich 1992]{FT92} Filippenko, A., \& Terlevich,
R. 1992, ApJ
         397, L79
\bibitem[Genzel et al. 1998]{GEN98} Genzel, R., Lutz, D., Sturm, E., Egami,
E., Kunze, D.,
         Moorwood, A.F.M., Rigopoulou, D., Spoon, H.W.W., Sternberg, A.,
Tacconi-Garman,
         L.E., Tacconi, L., \& Thatte, N. 1998, ApJ 498, 579
\bibitem[Gonzalez-Delgado \& P\'erez]{GDP} Gonzalez-Delgado, R., \& P\'erez,
E. 1997,
         APJS 108, 199
\bibitem[Ho, Filippenko \& Sargent 1993]{HFS93} Ho, L., Filippenko, A., \&
Sargent, W.
        1993, ApJ 417, 63
\bibitem[Kim, Veilleux \& Sanders 1998]{KVS} Kim, D.C., Veilleux, S., \&
Sanders, D.B.
         1998, ApJ 508, 627
\bibitem[Leitherer et al. 1999]{LEIT99} Leitherer, C., et al. 1999, ApJS
123, 3
\bibitem[Lutz et al. 1998]{LUT98} Lutz, D., Spoon, H.W.W., Rigopoulou, D.,
Moorwood,
         A.F.M., \& Genzel, R. 1998, ApJ 505, L103
\bibitem[Lutz, Veilleux \& Genzel 1999]{LVG} Lutz, D., Veilleux, S., \&
Genzel, R.
         1999, ApJ 517, L13
\bibitem[Melnick \& Mirabel 1990]{MM90} Melnick, J., \& Mirabel, I.F. 1990, A\&A 
         231, L19
\bibitem[Planesas, Colina \& P\'erez-Olea]{PCPO} Planesas, P., Colina, L.,
\& P\'erez-Olea, D.
        1997, A\&A 325, 81
\bibitem[Sanders]{SAN99} Sanders, D.B. 2000, {\it Proc. Conf. Space Infrared
Telescopes
 and Related Science}, 32nd COSPAR Meeting, eds. T Matsumoto \& T. de
Graauw, in press
\bibitem[Sanders \& Mirabel 1996]{SM} Sanders, D., \& Mirabel, I.F. 1996,
AR\&AA 34, 749
\bibitem[Scoville {\it{et al.}}~1999]{SCO99} Scoville, N.Z. et al. 2000,
         AJ, in press
\bibitem[Shields, G. 1990]{SHI90} Shields, G. 1990, ARAA 28, 525
\bibitem[Shields, J.C. 1992]{SH92} Shields, J.C. 1992, ApJ 399, L27
\bibitem[Surace et al. 1998]{SUR98} Surace, J.A., Sanders, D.B., Vacca,
W.D., Veilleux, S.,
         \& Mazzarella, J.M. 1998, ApJ 492, 116
\bibitem[Veilleux, Kim \& Sanders 1999]{VKS} Veilleux, S., Kim, D.C., \&
Sanders, D.B.
         1999, ApJ 522, 113
\bibitem[Veilleux, Sanders \& Kim 1999]{VSK} Veilleux, S., Sanders, D.B., \&
Kim, D.C.
         1999, ApJ 522, 139

\end{thebibliography}
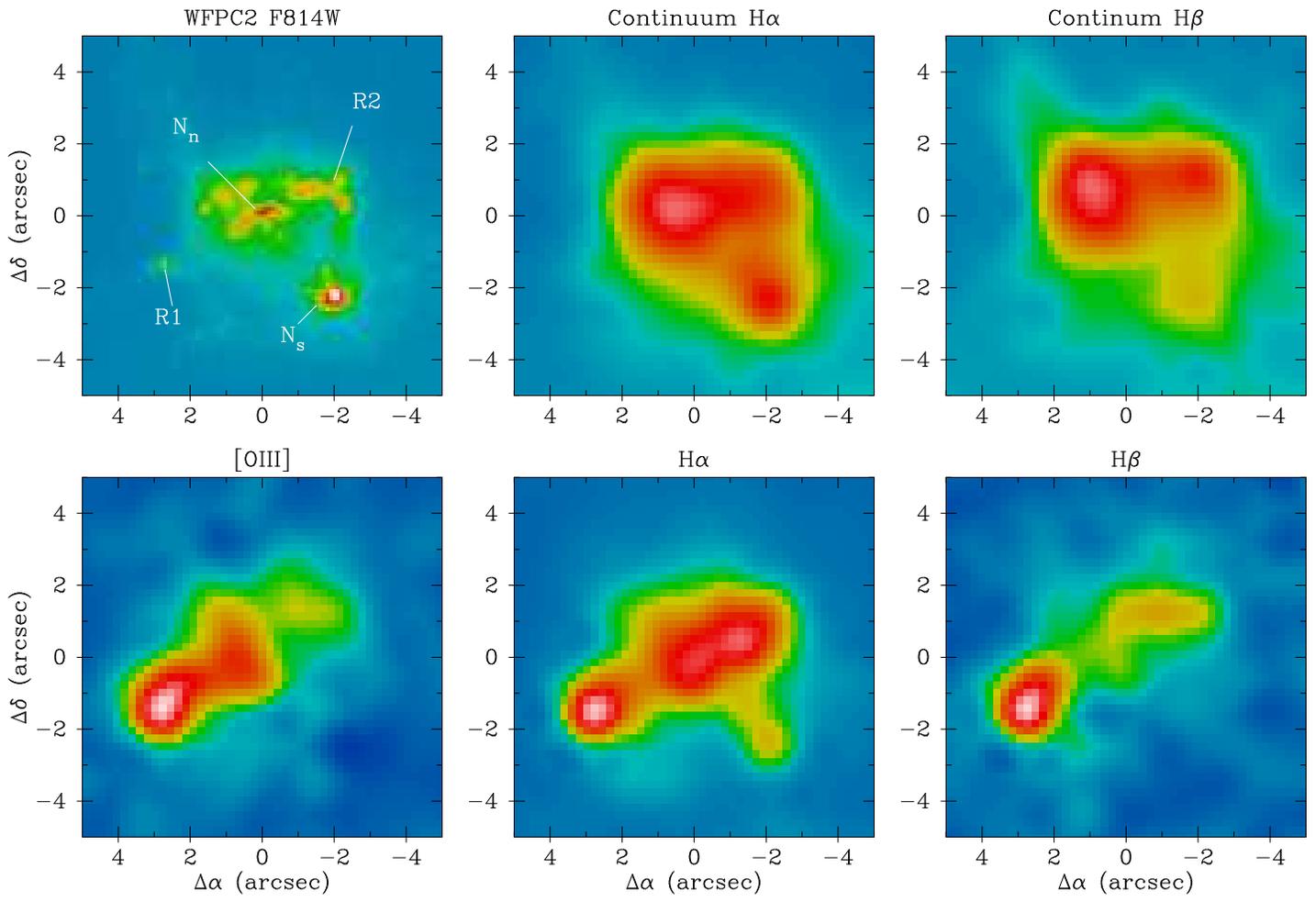
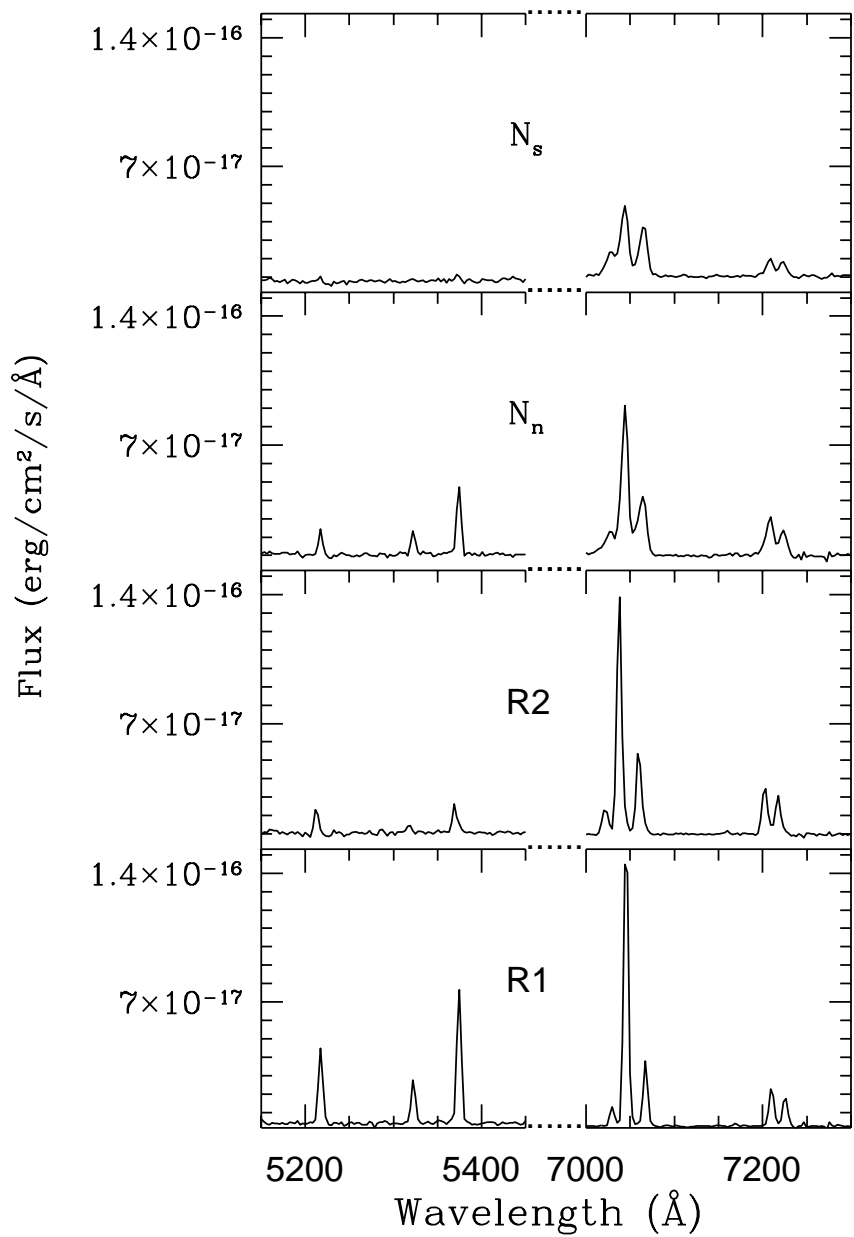
\end{document}